\documentclass[runningheads]{llncs}
\usepackage{graphicx}
\usepackage{todonotes} 

\usepackage{comment}
\usepackage{array}
\usepackage{ifthen}

\newboolean{auxinfo}
\setboolean{auxinfo}{true}   

\newboolean{showoutline}
\setboolean{showoutline}{true}

\newboolean{showcomment}

\setboolean{showcomment}{true} 

\usepackage[moderate]{savetrees}

\ifthenelse{\boolean{auxinfo}}
  {}
  {
    \setboolean{showoutline}{false}
    \setboolean{showcomment}{false}
  }

\usepackage{xcolor}
\usepackage{amssymb}

\ifthenelse{\boolean{showcomment}}
   {\newcommand{\nb}[2]{\fcolorbox{gray}{yellow}{\bfseries\sffamily\scriptsize#1}{\sf\small$\blacktriangleright${\em #2}$\blacktriangleleft$}}
   \newcommand{\working}[1]{\fcolorbox{gray}{yellow}{{\bf #1}\emph{\scriptsize---in progress---}}}
   \newcommand{\TBD}[1]{\fcolorbox{gray}{yellow}{{\bf #1}\textbf{TBD}}} 
  }
  {\newcommand{\nb}[2]{}{}
   \newcommand{\working}[1]{}
   \newcommand{\TBD}[1]{} 
  }

\usepackage{todonotes}
\ifthenelse{\boolean{showoutline}}{
	\newcommand{\outline}[3]{
		~\newline 
		\fcolorbox{red}{white}{
			\parbox{\columnwidth}{
				\ifthenelse{\equal{#1}{}}{
					\ifthenelse{\equal{#2}{}}{
						\noindent\colorbox[rgb]{0.65,0.16,0}{\textcolor[rgb]{1,1,1}{\textbf{Outline}}}
					}{
						\colorbox[rgb]{0.65,0.16,0}{\textcolor[rgb]{1,1,1}{\textbf{Outline -- Responsible: #2}}}
					}
				}{
					\ifthenelse{\equal{#2}{}}{
						\noindent\colorbox[rgb]{0.65,0.16,0}{\textcolor[rgb]{1,1,1}{\textbf{#1 page(s)}}}
					}{
						\colorbox[rgb]{0.65,0.16,0}{\textcolor[rgb]{1,1,1}{\textbf{#1 page(s) -- Responsible: #2}}}
					}
				}
				#3
			}
		}
	}
}{
	\newcommand{\outline}[3]{}
}

\newcommand\defauxcomm[1]{
       \expandafter\newcommand\csname #1\endcsname[1]{\nb{#1}{##1}}
       \expandafter\newcommand\csname WK#1\endcsname{\working{#1}}
       \expandafter\newcommand\csname TBD#1\endcsname{\nb{#1}}
    } 
   
\defauxcomm{KS}
\defauxcomm{MN}

\usepackage[normalem]{ulem}
\ifthenelse{\boolean{showcomment}}
    {\newcommand{\strike}[1]{\textcolor{red}{\sout{#1}}}}
    {\newcommand{\strike}[1]{}}

\begin{document}
%
\title{Between Policy and Practice: GenAI Adoption\\ in Agile Software Development Teams}
%
\titlerunning{GenAI Adoption in Agile Software Development Teams}
%
\author{Michael Neumann\inst{1}\orcidID{0000-0002-4220-9641} \and
Lasse Bischof\inst{1}\orcidID{0009-0002-6622-0770}\and
Nic Elias Hinz\inst{1}\orcidID{}\and
Abdullah Altun\inst{1}\orcidID{}\and
Luca Stockmann\inst{1}\orcidID{}\and
Dennis Schrader\inst{1}\orcidID{}\and
Ana Carolina Ahaus\inst{1}\orcidID{}\and
Erim Can Demirci\inst{1}\orcidID{}\and
Benjamin Gabel\inst{1}\orcidID{}\and
Maria Rauschenberger\inst{2}\orcidID{0000-0001-5722-576X}   \and
Philipp Diebold\inst{3,4}\orcidID{0000-0002-3910-7898} \and
Henning Fritzemeier\inst{5} \and
Adam Przybyłek\inst{6}\orcidID{0000-0002-8231-709X}}

\authorrunning{M. Neumann et al.}
%
\institute{University of Applied Sciences and Arts Hannover,\\ Hannover, Germany 
\email{michael.neumann@hs-hannover.de}\\
\and
University of Applied Sciences Emden/Leer,\\  Emden/Leer, Germany
\email{maria.rauschenberger@hs-emden-leer.de}\\
\and
Bagilstein GmbH, Mainz , Germany \email{philipp.diebold@bagilstein.de}
\and
IU International University, Erfurt, Germany \email{philipp.diebold@iu.org}
 \and
Volkswagen AG, Wolsburg, Germany \email{henning.fritzemeier@volkswagen.de}
 \and
University of Galway, Galway, Ireland \email{adam.przybylek@gmail.com}
}
\maketitle              
\begin{abstract}
\textit{Context:} The rapid emergence of generative AI (GenAI) tools has begun to reshape various software engineering activities. Yet, their adoption within agile environments remains underexplored.
\textit{Objective:} This study investigates how agile practitioners adopt GenAI tools in real-world organizational contexts, focusing on regulatory conditions, use cases, benefits, and barriers.
\textit{Method:} An exploratory multiple case study was conducted in three German organizations, involving 17 semi-structured interviews and document analysis. A cross-case thematic analysis was applied to identify GenAI adoption patterns.
\textit{Results:} Findings reveal that GenAI is primarily used for creative tasks, documentation, and code assistance. Benefits include efficiency gains and enhanced creativity, while barriers relate to data privacy, validation effort, and lack of governance. Using the Technology-Organization-Environment (TOE) framework, we find that these barriers stem from misalignments across the three dimensions. Regulatory pressures are often translated into policies without accounting for actual technological usage patterns or organizational constraints. This leads to systematic gaps between policy and practice.
\textit{Conclusion:} GenAI offers significant potential to augment agile roles but requires alignment across TOE dimensions, including clear policies, data protection measures, and user training to ensure responsible and effective integration.

\end{abstract}
\section{Introduction}
The rapid development and use of generative artificial intelligence (GenAI) in recent years has precipitated a profound transformation across a wide array of economic and technological domains. We see a high impact even in institutional contexts such as higher education~\cite{Neumann.2023a} or public administration~\cite{Salah.2023}. Also, the rise of GenAI since the release of ChatGPT in November 2022 has had various effects on our daily work and life in the various industries~\cite{Ebert.2023,Nguyen.2025a}. 

Today, GenAI tools such as ChatGPT or GitHub Copilot are used in a wide range of software engineering~\cite{Banh.2025} and thus led to a change in the way software development teams work and operate~\cite{Russo.2024}. For example, we know that software professionals apply GenAI for routine tasks like writing source code~\cite{Mendes.2024} or test cases~\cite{Arora.2024,Dong.2025}. Also, further approaches like AI agent-based tools are applied in practice to support the elicitation and analysis of requirements~\cite{Sami.2025,Zhang.2024}. We know from the growing research body that the effects of GenAI adoption in software development are diverse. Besides a growing performance of software development teams, other authors mention negative effects and risks~\cite{Sauvola.2024,Santos.2025}, \textit{e.g.,} on source code quality~\cite{Oertel.2025,Tanzil.2024}. Thus, the importance of human-integration is emphasized in the literature~\cite{Rauf.2025}, in particular, by providing specific values and principles in the Copenhagen Manifesto~\cite{Russo.2024a}. 

Within the area of agile software development (ASD), in particular, GenAI tools promise to automate routine tasks, foster creativity, and significantly improve the productivity of software developers and project stakeholders~\cite{Zhang.2025}. Agile methods, characterized by iterative development cycles, cross-functional collaboration, and continuous feedback, provide a particularly conducive environment for the piloting and integration
of GenAI tools into established workflows~\cite{Diebold.2025}. However, the deployment of these sophisticated systems also introduces a range of new challenges. According to Bahi et al. \cite{Bahi.2024}, practitioners are confronted with critical questions such as: “Is the use of such tools permissible?” and “Can sensitive data be safely disclosed to GenAI services?”. 

As agile methods are focusing strongly on social interaction and communication, one may assume that various agile related practices with a strong emphasis on social interaction such as pair programming are affected~\cite{Stray.2025a} when specific tasks are performed with GenAI support instead of humans~\cite{Alami.2025}. We know that the facet how agile software development teams specifically adopt GenAI tools in their practice is underrepresented in literature.

Thus, in this study, we address the following research questions:
\begin{itemize}
    \item\textbf{RQ 1:} What organizational and regulatory conditions shape GenAI adoption in practice?
    \item\textbf{RQ 2:} Which use cases of GenAI tools are adopted in agile software development teams?
    \item\textbf{RQ 3:} What benefits and barriers do agile team members associate while they adopt GenAI?
\end{itemize}

The paper is structured as follows: In the next Section, we provide an overview of the related work and present the contributions of our study in Section~\ref{sec:RelWork}. In Section~\ref{sec:ResearchDesign}, we explain how we applied the multiple case study. Next, we present our results and answer the research questions in Section~\ref{sec:Results}. 
The paper closes with a conclusion in Section~\ref{sec:Conclusion}.

\section{Literature Review}
\label{sec:RelWork}
In this Section, we provide an overview of the identified related work. As related work in the area of ASD is limited, we searched with a broader scope in the software engineering field and found some studies closely related to ours. 

The study most closely related to ours is by Kemell et al.~\cite{Kemell.2025}, who conducted a multiple case study across seven companies to examine GenAI adoption, identifying relevant use cases and challenges. They found six organizational barriers, including data privacy and legacy system concerns, as well as issues related to the black-box nature of LLMs. From a user perspective, they reported four further challenges, notably the complexity of learning to apply GenAI tools—particularly regarding prompt formulation in daily work.

Karlovs-Karlovskis~\cite{Karlovs.2024} supports these findings in his systematic literature review. Among 117 analyzed studies, about 65\% focus on GenAI-based code generation, revealing major gaps in empirical research on real-world integration and a lack of studies in other domains with potential for cost reduction, innovation, and optimization. Similar observations are made by Nguyen Duc et al.~\cite{Nguyen.2025a}, whose focus group covering eleven software engineering areas shows that research concentrates on implementation, quality assurance, and maintenance, while requirements engineering, design, and education remain underrepresented. They nonetheless argue that GenAI is likely to fundamentally transform the discipline, despite the early stage of current research.

Overall, existing research on GenAI in software engineering—especially within ASD—remains limited. While some studies (\textit{e.g.}, Ulfsnes et al.~\cite{Ulfsnes.2024}; Coutinho et al.~\cite{Coutinho.2024}) highlight productivity gains and tool versatility, they also point to persistent issues in knowledge sharing, reliability, and security. Bahi et al.~\cite{Bahi.2024} similarly show that GenAI can support Scrum practices but emphasize ongoing challenges across the development lifecycle, underscoring the paucity of empirical evidence from real-world agile environments.

Recent XP conference editions also hosted two workshops on GenAI and Agile (2024; 2025), producing exploratory papers on GenAI use in specific agile practices, such as Test-Driven Development~\cite{Mock.2025}, and broader topics like responsibility~\cite{Ulfsnes.2025} and teamwork effects~\cite{Kwok.2026}. While informative, their exploratory nature and workshop format limit their applicability to our research questions.

To address the identified gaps, this qualitative study examines the use of GenAI within agile software teams. It explores how GenAI is applied in practice, the organizational implications of its adoption, and relevant use cases along with associated benefits and barriers. By grounding the investigation in a real-world agile setting, this study provides empirical insights into how GenAI is reshaping software development dynamics in mid-sized enterprises.

\section{Research Design}
\label{sec:ResearchDesign}
Guided by Runeson and Höst’s guidelines~\cite{Runeson.2009}, this study adopts an exploratory qualitative multiple-case design to investigate how GenAI is embedded, applied, and experienced within the various roles operating in agile software-development teams. Case studies are well-suited for examining modern software engineering practices in real practical contexts, especially when the phenomenon cannot be easily separated from its context. For this study, the units of analysis are the employees operating in the agile software development teams. 

\subsection{Case Selection \& Context}
Our multiple case study was conducted in three German organizations. We selected the cases to cover a wide range of diverse aspects like roles, applied methods, experiences, and expertise.  All organizational information such as company, site, or team names had to be anonymized due to confidential reasons. Below we introduce the three companies:\\
\textbf{Case 1:} Dinoco is one of the world’s largest automobile manufacturers, operating more than nine vehicle brands and producing automobiles, motorcycles, and trucks worldwide. Our study took place at one of its German software development sites, the Dinoco Software Development Expertise Site (DSDES), which spans multiple locations and employs around 450 people. As part of Dinoco’s digitization strategy, DSDES develops strategic software products—such as cockpit software and over-the-air update processes—across more than ten agile teams. These teams follow eXtreme Programming principles within the Scaled Agile Framework (SAFe).

\begin{table}[ht]
\centering
\begin{tabular}{@{}>{\centering\arraybackslash}p{1cm}|
                >{\centering\arraybackslash}p{3cm}|
                >{\centering\arraybackslash}p{3cm}|
                >{\centering\arraybackslash}p{2.5cm}@{}}
\textbf{ID} & \textbf{Role} & \textbf{Exp. with Agile Methods} & \textbf{Case} \\ 
\hline
P01 & Agile Coach & 10 years & Dinoco \\
\hline
P02 & Product Owner & 13 years & Dinoco \\
\hline
P03 & Software Developer & 8 years & Dinoco \\
\hline
P04 & Software Developer & 10 years & Dinoco \\
\hline
P05 & Software Developer & 8 years & GMT \\
\hline
P06 & Software Developer & 30 years & GMT \\
\hline
P07 & Software Developer & 18 years & GMT \\
\hline
P08 & Product Owner & 15 years & GMT \\
\hline
P09 & Software Developer & 25 years & GMT \\
\hline
P10 & Product Owner & 11 years & GMT \\
\hline
P11 & Agile Coach & 5 years & Insight Inc. \\
\hline
P12 & Lean Coach & 10 years & Insight Inc. \\
\hline
P13 & Agile Coach & 10 years & Insight Inc. \\
\hline
P14 & Agile Coach & 5 years & Insight Inc. \\
\hline
P15 & Junior Agile Coach & 3 years & Insight Inc. \\
\hline
P16 & Agile Coach & 13 years & Insight Inc. \\
\hline
P17 & Agile Coach & 5 years & Insight Inc. \\
\hline
\end{tabular}
\vspace{0.5em}
\caption{Interviewee profiles}
\label{tab:participant-overview}
\end{table}
\vspace{-5mm}

\textbf{Case 2:} Gray Matter Technologies (GMT) is a medium-sized company in Germany and Austria with around 150 employees across three sites. GMT develops software products for external clients, including an ERP system and web shop applications. Its software development division consists of four agile teams, each dedicated to specific customer projects and working with an adapted Scrum approach. GMT’s clients range from large retail and manufacturing companies to medium-sized craft businesses.

\textbf{Case 3:} Insight Inc. is a small, fully remote consulting company in Germany with eight employees. It provides agile coaching and training and offers interim roles in which staff embed within client teams for six months to two years, typically as Scrum Masters and occasionally as Product Owners. Insight Inc.’s clients span various industries; for the employees involved in this study, the relevant clients were in the banking sector.

\subsection{Data Collection}
Runeson \& Höst~\cite{Runeson.2009} emphasize that case studies must combine direct, indirect, and independent data sources so that weaknesses inherent in  single method can be counterbalanced through data source triangulation. For our data collection we comprised mainly two data sources:

1) 17 semi-structured interviews through all three cases. A detailed overview including the interviewee profiles is given in Table~\ref{tab:participant-overview}. The interviews were conducted using a interview guideline which we created based on the Goal-Question-Metric (GQM) approach~\cite{Basili.1994}. The interview guide consist of four core clusters considering the personal background of the interviewee and the three main themes among the research questions: GenAI related contextual information, use cases of GenAI adoption, and barriers/benefits of the daily use of GenAI. The interview guide is available in our research protocol~\cite{Neumann.2025}. All interviews were held online using Microsoft Teams between March and May 2025 and have been recorded. The interviews lasted around 30 minutes for Dinoco, between 30 and 45 minutes for Insight Inc. and Gray Matter Technologies. To increase the quality, we performed a pre-test to verify the structure of the interview guide. The interviews were conducted by two researchers, while one researcher performed the interview based on the interview guide and the other researcher took an observing role during the interview and made notes. 

2) Document analysis: Besides the interviews, we gained information on analyzing internal documents, especially related to GenAI. Here, we covered internal GenAI guidelines for the use in software development processes. It is worth to mention that only Gray Matter Technologies provides a specific internal guideline for the GenAI use. The other cases provide a frame, which can be understood as a living artifact. 

\begin{figure*}[t]
\centering
\includegraphics[width=\textwidth]{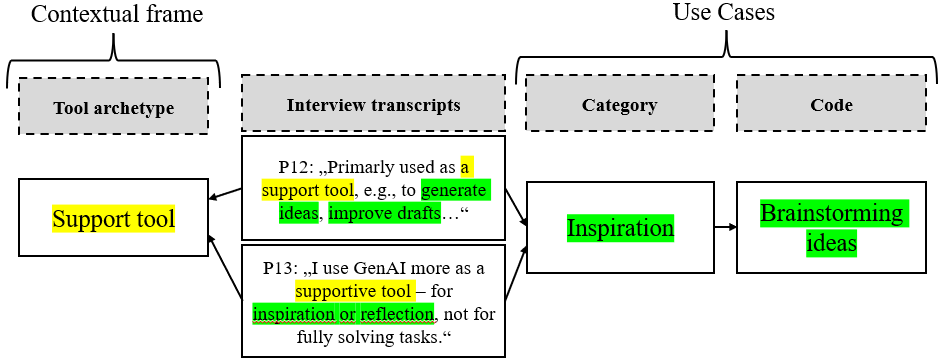}
\caption{Example of the coding process}
\label{fig:CodingExample}
\end{figure*}

\subsection{Data Extraction \& Analysis}

\textbf{Thematic Analysis \& Coding:} To analyze the qualitative data, we began by transcribing the recorded interviews using Microsoft Teams and Whisper AI. The resulting transcripts were subsequently reviewed and edited by the second and third authors to ensure accuracy and completeness. These validated transcripts formed the basis for a thematic analysis aimed at identifying, interpreting, and describing patterns within the qualitative material. We followed the procedural steps outlined by Braun and Clarke \cite{Braun_Clarke.2006}, which include familiarizing oneself with the data, generating initial codes, identifying themes, reviewing themes, and defining and naming them, followed by the reporting phase.

In order to establish a reliable and consistent coding process, multiple authors (authors 2–4 for the Dinoco case; 5–7 for GMT; and author 8 for Insight Inc.) independently coded a subset of the material. Their coding outputs were then compared and discussed with the first author. Any divergences were resolved through collaborative deliberation, fostering a shared understanding of the data and enhancing the stability of the thematic structure. This iterative procedure ensured that the resulting themes were both empirically grounded and aligned with the overall aims of the study.

\textbf{Cross-Case Analysis:} To strengthen the results of the multiple-case study, we conducted a cross-case analysis following Yin~\cite{Yin.2009}. For each case, the data were transferred into Microsoft Excel and examined for thematic overlaps, using separate sheets for contextual information, role-specific use cases, and identified barriers and benefits. We first compared context-related information (\textit{e.g.,} licensed GenAI tools or usage guidelines) across all three cases. Second, we analyzed the use cases, all of which appeared in at least two cases, meaning no borderline cases occurred. Third, we compared categories and codes from the interviews to identify overlaps in reported barriers and benefits. The analysis was carried out by the first author and validated by two additional researchers (the second and last author), with no discrepancies identified.

\subsection{Threats to Validity}
\label{sec:ThreatsToValidity}
\textbf{Construct validity:} Our interview participants may have interpreted specific terms related to GenAI (LLM or GPT) differently or assigned varying meanings to key concepts such as promoting strategies and "guidelines". To address this, we provided clarification and used "such as Github CoPilot" or asked participants for specific examples to clarify their understanding and ensure consistency across responses.

\textbf{Internal validity:}  We took specific actions to follow the same approach in conducting the interviews to mitigate potential bias. First, the interview guidelines was composed of neutral, non-leading questions. Additionally, the interviews were semi-structured, allowing us to explore topics in depth based on the interviewee’s responses. Each interview was conducted by two researchers, ensuring rigor. Furthermore, we used different data sources to strengthen the internal validity of our findings. 

\textbf{External validity:} 
The external validity of this study is limited by the fact that all three cases are German organizations operating under similar regulatory conditions. While the cases differ in size and structure, they do not represent the full range of software development environments. As with qualitative case studies, the findings support analytical rather than statistical generalization and may not directly transfer to other organizational or regulatory contexts.

\section{Results}
\label{sec:Results}

\subsection{Contextual Information on GenAI Adoption}
Here, we answer our first research question: \textit{What organizational and regulatory conditions shape GenAI adoption in practice?}

Organizational conditions cover both licensing of GenAI tools and training or coaching for professional use of GenAI in practice. Notably, none of the three cases offers training or coaching for GenAI usage in their software development teams. Interestingly, at GMT, while company policy mandates at least two annual training sessions on AI applications and legal frameworks, interviewees either were unaware of these offerings or reported that the training focused exclusively on explaining the AI guideline itself rather than developing practical AI usage skills. This suggests a substantial gap between formal policy and practitioner needs. 

\begin{table}[h!]
\centering
\caption{Overview of licensed GenAI tools}
\begin{tabular}{lccc}
\hline
\textbf{Tool} & \textbf{Dinoco} & \textbf{GMT} & \textbf{Insight Inc.} \\ 
\hline
ChatGPT & no & no & yes \\
ChatGPT Pro Version & no & no & yes \\
GitHub Co-Pilot & yes & yes & no \\
Microsoft Co-Pilot & yes & yes & no \\
\hline
\end{tabular}
\label{tab_ai_tools}
\end{table}
\vspace{-5mm}

Regarding licensing, an overview is provided in Table~\ref{tab_ai_tools}. The governance mechanisms for tool selection differ markedly across cases. At Dinoco, licensing decisions are led by compliance and top management, following a risk-averse strategy that favors established vendors; for instance, Co-Pilot was licensed through existing Office 365 agreements. GMT, by contrast, follows a formal company guideline that classifies tools according to the EU AI Act’s risk pyramid (Minimal, Limited, High Risk)~\cite{EUAIACT.2024}, maintains an official whitelist, requires management approval, and involves an internal expert group for validating newly requested tools. Key criteria include customer data protection, legal compliance, data-leakage prevention, and control over tools transmitting usage data externally. At Insight Inc., no comparable governance structure exists.

Our findings also show that tools such as DeepL Write, Gemini, or NotebookLM are not licensed in any of the three cases. A gap exists between sanctioned tools and those used in practice. At both Dinoco and GMT, ChatGPT functions as a form of “shadow IT,” despite lacking approval. At Dinoco, this is enabled by an informal agreement that developers use ChatGPT in their “private time.” At GMT, survey and interview data confirm ChatGPT as the most frequently used tool. In contrast, Insight Inc. officially licenses ChatGPT, though employees occasionally rely on private accounts for sensitive topics to ensure data separation.

With regard to regulatory aspects, internal guidelines for the use of GenAI tools are of high importance. In the Dinoco case, the use of ChatGPT is not explicitly regulated, whereas clear policies exist for GitHub Co-Pilot and Microsoft Co-Pilot. Nevertheless, several interviewees reported using the Pro version of ChatGPT for work-related tasks. The specific use of these tools varies considerably across roles: while Agile Coaches and Product Owners use ChatGPT frequently, developers rely on it less often. Due to the integration of Co-Pilot tools into the IDE, their use is more seamless and compliance can be more easily monitored. Furthermore, regulatory aspects also include data protection, which is explicitly governed through agreements between Dinoco and the respective tool providers. For instance, P03 stated: ``Dinoco [...] \textit{has made an agreement with Copilot that the data may not be used to evaluate or train the model}." In the GMT case, several interviewees highlighted the risk that certain GenAI tools or models may not ensure sufficient data sovereignty, as user inputs and the information they contain can be used for model training. Given the opaque nature of AI model operations, it cannot always be determined how such data are further processed. Consequently, GenAI tools cannot be applied to code segments involving sensitive information, such as encryption keys or personal data. The absence of formal guidelines leads to users ``[...] \textit{often rely on personal judgement, experience or informal principles}"; according to P11. While this approach allows for flexibility, it also entails risks related to consistency, legal certainty, and accountability. P14 emphasized that formal frameworks are frequently missing within organizations, resulting in uncertainty, inconsistent implementation, and highly individualized practices. At the same time, there is a strong demand for clear, pragmatic and legally sound guidance. Such guidelines are particularly essential for fostering acceptance of the technology among less experienced users.

Concluding our findings for RQ1, we observe that while ChatGPT emerges as the most frequently used tool across all three cases, the organizational conditions governing GenAI adoption differ considerably. Licensing arrangements vary, with Insight Inc. providing company-licensed ChatGPT accounts while Dinoco and GMT license Co-Pilot tools instead. Governance mechanisms range from formalized approval processes with internal expert groups (GMT) to compliance-driven top-level decisions (Dinoco) to largely absent formal frameworks (Insight Inc.). Notably, even where policies exist, adherence is inconsistent; some employees circumvent regulations by using private accounts for work-related tasks.

\subsection{Applied GenAI Use-Cases in Agile Software Development}
Here, we answer our second research question: \textit{Which use cases of GenAI tools are adopted agile software development teams?}

Analyzing our findings during the cross-case analysis, we identified three main clusters of use cases and three roles applying in total 21 use cases. We categorized these 21 use cases thematically, identifying 6 as creative tasks, 7 as documentation-related tasks, and 8 as coding-related tasks.

Use case adoption varies substantially by role and experience level. Product Owners and Agile Coaches reported higher-intensity usage, particularly for text-based creative tasks, while developers' adoption ranged from minimal to moderate depending on codebase characteristics. At GMT, one senior developer with 15 years' experience self-rated usage intensity at 1 out of 4, attributing low adoption to established workflows and extensive domain knowledge that reduced perceived need for AI assistance. Conversely, newer team members and those in product-focused roles reported more extensive integration of GenAI into daily work.

\begin{figure}[t]
\centering
\includegraphics[width=\textwidth]{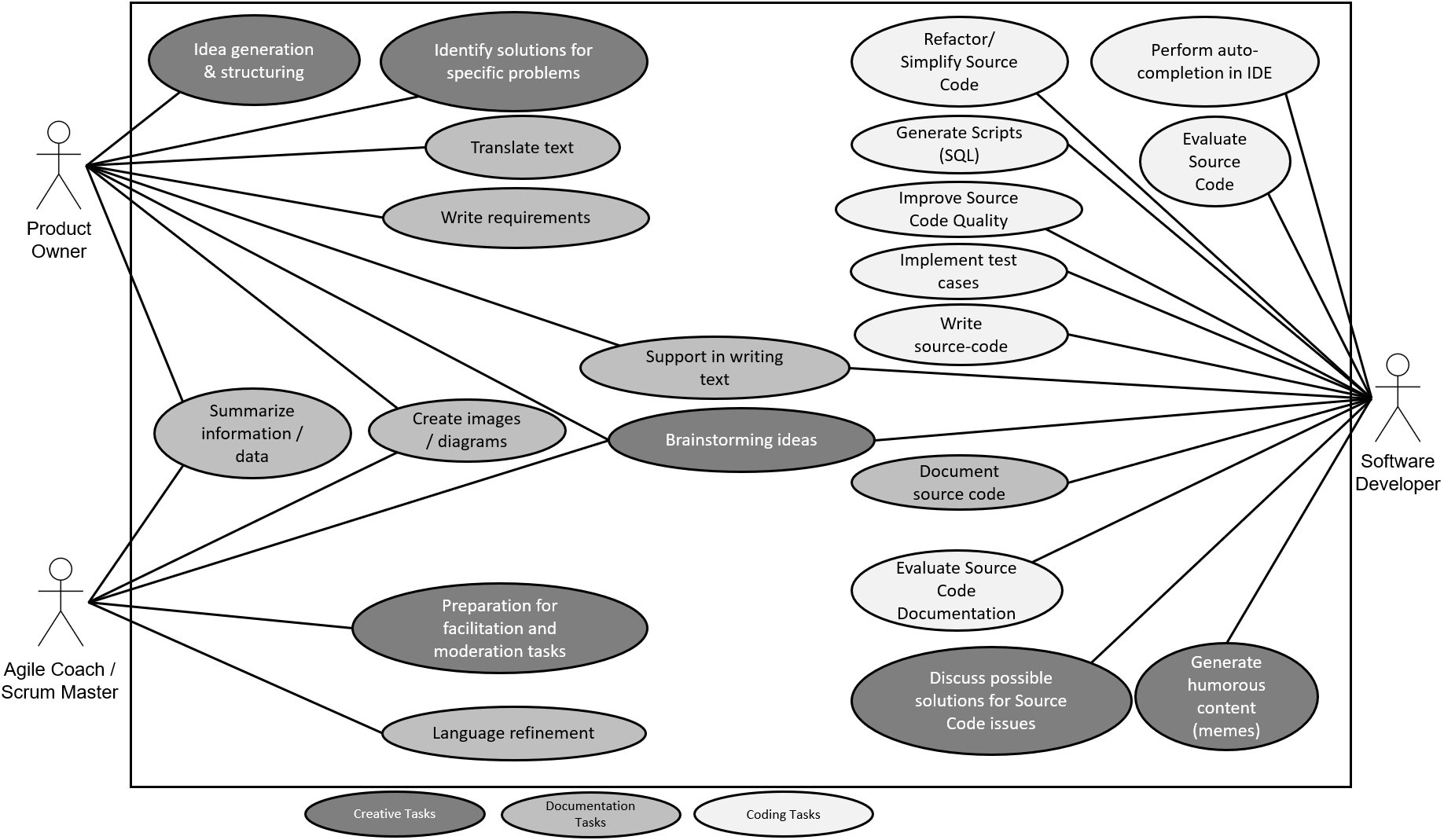}
\caption{Visualization of the identified use cases}
\label{fig:UseCases}
\end{figure}

Figure\ref{fig:UseCases} depicts the use cases applied by the various roles, to ensure a systematic description, we used Unified Modeling Language (UML) Use Case Analysis modeling for visualization. Furthermore, the result of the cross-case analysis including a detailed overview of the identified use cases per role and case is given in our research protocol~\cite{Neumann.2025}. In the research protocol, we also provide specific descriptions per use case applying a systematic schema due to space limitations in this paper.  

Across all three organizations, GenAI is primarily leveraged as a personal productivity assistant, augmenting individual tasks rather than transforming team collaboration. As shown in Figure~\ref{fig:UseCases}, we identified a clear pattern of role-specific adoption, with Product Owners, Agile Coaches/Scrum Masters, and Developers applying the tools to distinct aspects of their work. We describe these patterns below using three archetypes: the Research Assistant for creative and conceptual tasks, the Virtual Tutor for documentation and communication, and the Pair Software Developer for coding and technical activities.

\textbf{Applied as a Research assistant:} At Dinoco, GenAI is used specifically to support creative processes. P02 uses GenAI tools for brainstorming and to scrutinize ideas: ``\textit{A lot for brainstorming, to have a practical communication partner and to validate a few ideas}." P1 also reports typical use cases in which GenAI helps to structure thoughts: ´´\textit{But then you often have these thought blocks: How do I get this into sensible words now}?". In these use cases, GenAI serves less as a pure source of information, but is seen as a discursive partner that provides impulses and removes creative blocks. A particularly prevalent use case across all roles and organizations is GenAI as an enhanced search mechanism, effectively replacing traditional web searches and Stack Overflow queries. Multiple developers characterized ChatGPT as their ``first point of contact'' for technical questions, valuing consolidated answers over navigating multiple search results. 

\textbf{Applied as a Virtual tutor:} Product owners in particular use GenAI
to formulate requirements more quickly. P01 describes this succinctly:
´´[...] \textit{writing features and stories}". This example shows that GenAI
is primarily used here to elaborate the content of user stories and thus simplifies
writing work. From an Agile Coach/Scrum Master perspective, tools like ChatGPT were frequently mentioned as primary assistants for content generation, including drafting texts, brainstorming ideas, and refining formulations. P12 described GenAI as a ´´\textit{very helpful sparring partner}” when initiating tasks or formulating complex content. P11 noted that while the tool does not necessarily
complete tasks for them, it significantly accelerates their work by ´´\textit{kicking off ideas faster}”. Furthermore, several Agile Coaches/Scrum Masters (P01, P12, P14) mentioned that they use GenAI to prepare for retrospectives by generating creative formats or by refining formulations in team surveys. Others highlighted that it helps them organize thoughts when faced with a blank page, particularly for internal presentations or training materials (e.g., P13). These scenarios suggest that GenAI is mainly used to support the early phases of
cognitive work—ideation, structuring, and polishing—rather than full automation. Beyond agile artifacts, Product Owners and coaches employ GenAI extensively for stakeholder communication. At GMT, one Product Owner (P8) reported using Microsoft Copilot daily to formulate customer emails, particularly for business English communication where uncertainty about professional phrasing exists. The tool serves as both translator and style consultant. At Dinoco, similar patterns emerge for refining internal communications and ensuring professional tone in external correspondence. This communication support function operates largely invisibly—improving message quality without transforming the underlying work process.

\textbf{Applied as a Pair Software Developer:} From a software development perspective, GenAI primarily plays a role in simplifying standard tasks. 
Its most valued contributions include generating ``\textit{boilerplate code}'' (P03), writing repetitive unit tests (P03), and providing ´´\textit{auto-completion within the IDE}'' (P04). Beyond generating new code, a common use case is understanding existing code; developers frequently ask the AI to explain complex snippets, especially when working in large, legacy codebases. Most interviewees reported using GenAI as a form of pair programmer to improve, refactor, or get a ´´\textit{second opinion}'' on code segments (P04). This interaction sometimes evolves into a collaborative dialogue to solve a problem. Documentation, a notoriously time-consuming task, is another key application, used for generating both inline comments and external commit messages. A developer at GMT noted they document more now simply because the AI makes it faster. These statements confirm that GenAI's value in development lies in automating routine work and augmenting the developer's understanding, allowing them to focus on more complex problem-solving.

Interestingly, none of the interviewees reported using GenAI for decision-making or for
directly facilitating meetings, citing concerns about quality, ethics, and team acceptance.
Instead, this technology is seen as a background tool that empowers but does not replace
the human element of agile ways of working. Overall, GenAI tools appear to be most effective when they are seamlessly integrated into existing processes and support specific work tasks. Their actual added value largely depends on the user’s role, the specific use case, and the individual team members’ approach to the technology.

\subsection{Benefits \& Barriers of GenAI Adoption}
In this subsection, we answer our third research question: \textit{What benefits and barriers do agile team members associate while they adopt GenAI?}

The benefits of GenAI use can be grouped into three main categories. First, GenAI enables efficiency gains in coding by simplifying or optimizing code segments through access to a far broader knowledge base than individual developers possess. Developers reported reduced time for boilerplate code generation, faster prototyping of solutions for customer requests, and accelerated resolution of unfamiliar technical challenges. GMT developers also reported learning new language features, design patterns, or optimization techniques from AI suggestions. Second, GenAI enhances overall task efficiency by automating routine activities and generating code skeletons, thereby accelerating development processes. It can also handle tedious tasks such as documentation, allowing developers to focus more on value-adding activities. Beyond efficiency gains, GenAI tools also foster creativity, particularly in early project phases such as requirements definition and concept development. They provide inspiration, structure, and support out-of-the-box thinking. Some interviewees also mentioned the use of prompt templates to streamline their work. These insights suggest that GenAI contributes not only to implementation but also to ideation and conceptual design.
Perceived benefits, however, vary by role. While Product Owners and Requirements Engineers described GenAI as a valuable support tool, Software Developers expressed more reserved attitudes, indicating that the impact of GenAI depends strongly on professional role and context of use. This divergence is particularly pronounced at Dinoco, where developers characterized productivity gains as ´´\textit{modest or even neutral}'', with one senior developer (P04) asserting that ´´\textit{productivity cannot meaningfully be measured}''—calling into question the empirical basis for claimed efficiency improvements.

The identified barriers can be grouped into three main clusters. The first concerns the effort required to validate GenAI outputs and the risk of uncritically adopting generated results. At GMT, this validation burden is formalized through the AI guideline, which mandates that AI-generated content may only be used after two independent, authorized persons have reviewed and approved it via email. However, this rule is unworkable in practice, as the overhead of a two-person review for every AI-generated snippet would create unsustainable bottlenecks.

Several interviewees perceived this as problematic, noting that hidden code dependencies demand a thorough understanding of functionality for correct operation and debugging. As GenAI occasionally produces inaccurate outputs, critical evaluation remains essential.
The second cluster relates to the effort involved in effectively using GenAI. Interviewees emphasized that meaningful results often require substantial input and context, yet the outcome may still lack value. Developers at Dinoco stressed that supplying sufficient context is time-consuming, and incomplete prompts frequently lead to generic or unusable responses. Enhancing usability—by simplifying interaction and reducing reliance on prompt engineering—was seen as crucial. Better integration into existing tools, such as Copilot in Visual Studio Code, was cited as an effective solution, as it allows suggestions directly within the IDE. Beyond the practical effort of using GenAI, several interviewees highlighted bureaucratic challenges, primarily stemming from GMT's AI guidelines. The mandatory evaluation of GenAI models prior to use was perceived as a barrier to experimentation. Moreover, employees at Dinoco and GMT are responsible for managing their own accounts, which requires identifying suitable models and verifying their approval status. If a model has not yet been evaluated, an assessment must be requested or an approved alternative located. Some interviewees also expressed concern that GenAI use might prompt uncomfortable questions or require justification to supervisors, particularly when no tangible results are achieved. For larger organizations a central obstacle lies in the organizational restrictive attitude towards GenAI tools. 
This was addressed by several respondents. P01 emphasized: ´´\textit{My experience with} [Dinoco] \textit{- or with other companies - is that everything is always so restrictive and hostile to innovation}." This assessment points to structural hurdles within large organizations that can slow down innovation due to safety concerns or bureaucratic processes. At Dinoco, the compliance department's approval requirement acts as a bottleneck; ChatGPT's non-approved status leads developers to use it anyway, reframed as 'private time' usage. Although there are official regulations on the use of GenAI, the actual implementation is often left to the employees themselves. Here it becomes clear that the responsibility for compliant behavior is highly individualized, while control or assistance from the company is only available to a limited extent. The barrier is thus not merely bureaucratic friction, but the lack of organizational pathways to reconcile security requirements with practical needs—forcing employees to choose between compliance and productivity.

In summary, we can conclude, that the interviewees reported time savings, enhanced creativity, and support for routine tasks as major benefits. However, they also expressed concerns about data privacy,
dependency on AI-generated output, and a lack of transparency. These mixed perceptions underscore the importance of context-sensitive implementation and support mechanisms.

\section{Discussion}
\label{sec:Implications}
Our findings contribute to the growing body of empirical research on GenAI adoption in software development by offering a practice-centered perspective from within agile teams. To interpret the multifaceted nature of this adoption, we analyze our results through the lens of the Technology-Organization-Environment (TOE) framework~\cite{Depietro.1990}, examining how technological characteristics, organizational structures, and environmental contexts shape the integration of GenAI.

\textbf{Technology Context: } The technological dimension centers on tool capabilities and limitations. In line with Kemell et al.~\cite{Kemell.2025}, we observe that GenAI is predominantly used as a personal productivity assistant, augmenting individual performance rather than fundamentally altering team-level processes.  Recent studies by Russo~\cite{Russo.2024} and Banh et al.\cite{Banh.2025} highlight the importance of seamless workflow integration as a prerequisite for sustained GenAI adoption. Our findings corroborate this, demonstrating that tools embedded into existing development environments, such as Copilot in VS Code, facilitate coding tasks with minimal workflow disruption. However, the technological context also reveals critical limitations. The validation burden associated with AI-generated outputs represents a fundamental technological constraint—current GenAI tools cannot guarantee accuracy, necessitating human review that partially offsets efficiency gains. Interviewees emphasized that meaningful results require substantial context, yet supplying it is time-consuming, and incomplete prompts frequently lead to generic or unusable responses.

Our results further indicate that perceived benefits vary substantially by experience level and organizational context. Senior developers working with legacy systems reported minimal productivity gains, challenging universal efficiency narratives. This suggests that GenAI's value is contingent on factors including domain expertise, codebase maturity, and established workflow efficiency rather than being universally transformative.

\textbf{Organization Context: } The organizational dimension encompasses role structures, internal capabilities, and the cultural readiness of teams to adopt GenAI. By focusing explicitly on agile roles, our study adds depth to existing insights: GenAI is not only integrated into individual developer workflows but also into the preparatory routines of Product Owners and Scrum Masters. This role-specific appropriation reflects how organizational structures shape technology use patterns.
While Kemell et al.~\cite{Kemell.2025} emphasize organizational conditions and governance mechanisms, our findings highlight the relevance of role-specific affordances and agile values in shaping GenAI usage. Developers, for instance, leverage GenAI to offload repetitive coding tasks, whereas Scrum Masters use it to ideate formats for team reflection. The alignment between GenAI affordances and role-specific task demands determines adoption intensity and perceived value.
Crucially, teams appropriate GenAI to align with agile values. Rather than automating core processes, the technology supports—but does not substitute—activities like requirement elicitation and retrospective design. This subtly reshapes work to amplify efficiency while preserving underlying social dynamics. However, organizational readiness remains limited; the absence of adequate training across all cases forces practitioners to develop competencies independently, creating a significant capability gap.

\textbf{Environment Context: }The environmental dimension captures external regulatory pressures, industry norms, and competitive dynamics that shape GenAI adoption. Our findings reveal that data protection requirements constitute the most significant environmental constraints on adoption.
Across all roles, GenAI remains subordinate to human judgment, reflecting both epistemic caution and adherence to human accountability. This is not merely a practitioner preference but, in some cases, an institutional mandate—at GMT, internal policies explicitly frame AI outputs as ``recommendations'' requiring human review, institutionalizing a human-in-the-loop practice. These institutional requirements reflect broader environmental pressures stemming from the EU AI Act's risk classification framework, data protection regulations (GDPR), and sector-specific compliance mandates. While this formalized response creates structured pathways for adoption, it simultaneously introduces bureaucratic friction.

\textbf{Cross-Dimensional Tensions and the Compliance Gap: } A central contribution of our study is identifying systematic tensions across TOE dimensions that manifest as compliance gaps. Viewed through the TOE lens, this gap arises when environmental pressures (regulatory compliance) are translated into organizational policies without adequate consideration of technological characteristics and organizational constraints.

Our analysis reveals that a misalignment between top-down governance and bottom-up work practices creates systemic friction. When organizational policies, designed to mitigate risk, fail to account for the day-to-day productivity needs of practitioners, they are often perceived as impractical and are informally bypassed. This policy-practice gap manifests in distinct ways across our cases, but consistently leads to shadow IT practices. For instance, GMT’s mandatory two-person review for all AI outputs illustrates how strict compliance can conflict with the need for agile velocity. At Dinoco, developers reframe unapproved ChatGPT usage as ``private time''—an adaptive response that bypasses official regulations to resolve the conflict between policy and productivity. Furthermore, even where organizations provide licensed tools, we found that data protection constraints prohibiting realistic work content can render them impractical, driving practitioners toward unlicensed alternatives for convenience.

Ultimately, when tools offer substantial productivity benefits, prohibition without practical alternatives merely shifts usage into informal arrangements that may increase organizational risk. This challenges the assumption that restrictive policies alone ensure compliance, suggesting that effective governance requires creating frameworks practitioners can and will actually follow.

\section{Conclusion \& Future Work}
\label{sec:Conclusion}
This multiple-case study investigates how agile practitioners currently employ GenAI tools in real organizational settings. The findings show that GenAI is already integrated into a wide range of agile practices—from repetitive tasks such as code implementation and user-story writing to more strategic activities like team facilitation and reflection. Yet its adoption also presents challenges, including legal ambiguities, the need for governance frameworks, and concerns about data security and trust. Overall, while GenAI offers considerable potential to strengthen agile roles, its integration must be carefully managed to align with organizational structures and policies.

Although software development teams in several organizations already use GenAI, this raises further questions: Do other organizations employ GenAI in similar ways, and does access to adequate training improve team effectiveness? Addressing these questions requires more extensive research that compares current practices across the broader industry. Future studies could, for example, examine additional organizations with software development teams and contrast their approaches and outcomes.

\section*{Acknowledgements}
This publication has emanated from research jointly funded by Taighde Éireann – Research Ireland under Grant Number 13/RC/2094\_2, and co-funded by the European Union under the Systems, Methods, Context (SyMeCo) programme Grant Agreement Number 101081459. Views and opinions expressed are however those of the author(s) only and do not necessarily reflect those of the European Union or the European Research Executive Agency. Neither the European Union nor the granting authority can be held responsible for them.

 \bibliographystyle{splncs04}
 \bibliography{references}

\end{document}